\documentclass[a4paper]{jpconf}
\usepackage{graphicx}
\usepackage{amsfonts}
\usepackage{amsmath}
\usepackage{cite}
\usepackage{color}
\usepackage{tikz}
\usetikzlibrary{matrix}
\usepackage{float}

\newcommand\Mvec{\,\mbox{\bf M}}
\newcommand\Li{\,{\rm Li}}
\newcommand\HH{\,{\rm H}}
\newcommand\Ti{\,{\rm Ti}}
\newcommand\w{\,{\sf w}}
\newcommand{\ds}{\displaystyle}
\newcommand\SH{\,\mbox{$\sqcup \! \sqcup$}\,}
\renewcommand{\H}[2]{\textnormal{H}_{#1}\hspace{-0.2em}\left(#2\right)}
\renewcommand{\S}[2]{\textnormal{S}_{#1}\hspace{-0.2em}\left(#2\right)}

\begin{document}

\title{{\footnotesize{\sf DESY 13-189, DO-TH 13/28, SFB/CPP-13-78, LPN13-077}}\\
Generalized Harmonic, Cyclotomic, and Binomial Sums, their Polylogarithms and Special Numbers
}

\author{J Ablinger$^a$, J  Bl\"umlein$^b$\footnote{Speaker}, and C Schneider$^a$}

\address{$^a$ Research Institute for Symbolic Computation (RISC)
Johannes Kepler University, \hspace*{1.5mm} Altenbergerstra\ss{}e 69, A-4040 Linz, Austria\\
$^b$ Deutsches Elektronen-Synchrotron, DESY, Platanenallee 6, D-15738
Zeuthen, Germany}

\ead{Johannes.Bluemlein@desy.de}

\begin{abstract}
A survey is given on mathematical structures which emerge in multi-loop Feynman diagrams.
These are multiply nested sums, and, associated to them by an inverse Mellin transform, specific 
iterated integrals. Both classes lead to sets of special numbers. Starting with harmonic sums and
polylogarithms we discuss recent extensions of these quantities as cyclotomic, generalized (cyclotomic), 
and binomially weighted sums, associated iterated integrals and special
constants and their relations.
\end{abstract}

%%%%%%%%%%%%%%%%%%%%%%%%%%%%%%%%%%%%%%%%%%%%%%%%%%%%%%%%%%%%%%%%%%%%%%%%%%%%%%%%%%%%%%%%%%%
\section{Introduction}
%%%%%%%%%%%%%%%%%%%%%%%%%%%%%%%%%%%%%%%%%%%%%%%%%%%%%%%%%%%%%%%%%%%%%%%%%%%%%%%%%%%%%%%%%%%

\vspace*{1mm}
\noindent
During the late 1990ies several massless and massive two-loop calculations in Quantum 
Chromodynamics reached a complexity, see e.g. \cite{ZN1,ZN2,ZN3,ZN4,Blumlein:2012bf}, which made it 
necessary to introduce new functions a systematic manner to represent the analytic results 
in an adequate form. Dilogarithms, polylogarithms 
\cite{DILOG1,DILOG2,DILOG3,DILOG4,DILOG5,DILOG6,DILOG7,SPENCE,POLYLOG1,POLYLOG3,POLYLOG3a} 
and Nielsen integrals \cite{NIELSEN1,NIELSEN2,NIELSEN3} 
with complicated arguments did not allow to perform further calculations. Due to this harmonic sums, resp. 
specific types of Mellin transforms, were independently introduced in \cite{Vermaseren:1998uu} and 
\cite{Blumlein:1998if} as the basic building blocks. Shortly after the harmonic polylogarithms over 
the alphabet 
$\{1/x, 1/(1-x), 1/(1+x)\}$ were found \cite{Remiddi:1999ew}. These are iterated integrals of 
the Volterra-type having been studied by Poincar\'{e} \cite{POINC1,POINC2,POINC3,POLYLOG2} more than 
100 years before.

Physics expressions, such as massless and massive Wilson coefficients in the asymptotic region to 
2-loops, can be 
expressed in terms of harmonic sums only \cite{TWOLOOP1,TWOLOOP2,TWOLOOP3,TWOLOOP4,TWOLOOP5}. This 
also applies to the 3-loop anomalous dimensions 
\cite{ANDIM31,ANDIM32,Ablinger:2010ty,Blumlein:2012vq}. However, in the calculation of the 
massless 3-loop Wilson coefficients in deep-inelastic scattering \cite{Vermaseren:2005qc}, resp. the
massive case \cite{Ablinger:2010ty,Blumlein:2012vq,Ablinger:2012qm}, generalized harmonic sums, also 
called S-sums \cite{Moch:2001zr,Ablinger:2013cf}, emerge at least in intermediate results. For 
massive 3-loop graphs $4th$ and $6th$ root of unity weights contribute. At the side of the nested 
sums they belong to the cyclotomic harmonic sums \cite{Ablinger:2011te}. Furthermore, root-valued 
letters occur in the alphabets of iterated integrals \cite{ABRSW13}. They correspond to 
binomially-weighted generalized cyclotomic sums. Finally, also elliptic integrals emerge in the 
calculation of massive Feynman diagrams 
\cite{ELLIPTIC1,ELLIPTIC2,ELLIPTIC3,ELLIPTIC4,ELLIPTIC5,ELLIPTIC6}. Special numbers are associated to 
the above nested sums in the limit $N \rightarrow \infty$ and the iterated integrals for $x=1$. In the 
simplest case these are the multiple zeta values (MZVs) \cite{Blumlein:2009cf}. 
It is obvious, that more structures are expected to contribute calculating Feynman diagrams at even 
higher loops and for more legs. In this article we give a brief survey on the structures having been
found so far.\footnote{For a recent review see \cite{Ablinger:2013jta}.} The {\it method} to unravel 
these structures consists in applying modern summation techniques to the multiply nested sums, 
through 
which the corresponding Feynman diagrams are represented, and to solve them in difference fields 
using the algorithms \cite{SIG1,SIG2,SIG3,SIG4,SIG5,SIG6,SIG7,SIG8,SIG9} being encoded 
in the {\tt Mathematica}-package {\tt Sigma} \cite{SIG10,SIG11}. 
%%%%%%%%%%%%%%%%%%%%%%%%%%%%%%%%%%%%%%%%%%%%%%%%%%%%%%%%%%%%%%%%%%%%%%%%%%%%%%%%%%%%%%%%%%%
\section{Harmonic Sums, Polylogarithms and Multiple Zeta Values}
%%%%%%%%%%%%%%%%%%%%%%%%%%%%%%%%%%%%%%%%%%%%%%%%%%%%%%%%%%%%%%%%%%%%%%%%%%%%%%%%%%%%%%%%%%%

\vspace*{1mm}
\noindent
Let us consider the 2-point functions in Quantum Chromodynamics with local operator insertions.
Already in case of the quark-quark anomalous dimension the most simple harmonic sum
%----------------------------------------------------------------------------------------------
\begin{eqnarray}
S_1(N) = \sum_{k=1}^N \frac{1}{k},~~~~N \in \mathbb{N}
\end{eqnarray}
%----------------------------------------------------------------------------------------------
occurs, cf. e.g. \cite{Blumlein:2012bf}. At higher orders more general harmonic sums contribute.
They are defined by \cite{Vermaseren:1998uu,Blumlein:1998if}
%----------------------------------------------------------------------------------------------
\begin{eqnarray}
S_{b,\vec{a}}(N) = \sum_{k=1}^N \frac{({\rm sign}(b))^k}{k^{|b|}} S_{\vec{a}}(k),~~~S_\emptyset = 1
~~~~b, a_i \in \mathbb{Z} \backslash \{0\}~.
\end{eqnarray}
%----------------------------------------------------------------------------------------------
The Mellin transformation \cite{MELLIN1,MELLIN2} 
%----------------------------------------------------------------------------------------------
\begin{eqnarray}
\label{EQmel}
\Mvec[f(x)](N) = \int_0^1 dx~x^N~f(x)
\end{eqnarray}
%----------------------------------------------------------------------------------------------
relates harmonic sums to harmonic polylogarithms
%----------------------------------------------------------------------------------------------
\begin{eqnarray}
\label{EQ1}
S_{\vec{a}}(N) = \sum_c r_c \Mvec[\HH_{\vec{b}}(x)](N) + \sum_d r_d \sigma_{\vec{c}_d},~~~~r_c \in 
\mathbb{Q}~,
\end{eqnarray}
%----------------------------------------------------------------------------------------------
with $\sigma_{\vec{c}_d}$ multiple zeta values.
The harmonic polylogarithms \cite{Remiddi:1999ew} may be  defined as iterated integrals
%----------------------------------------------------------------------------------------------
\begin{eqnarray}
\HH_{b,\vec{a}}(x) &=& \int_0^x \frac{dy}{y - b} \HH_{\vec{a}}(y),~~~~b \in 
\{0,1,-1\},~~~~\HH_\emptyset(x) = 1 \\
\HH_{\underbrace{\mbox{\scriptsize 0, \ldots ,0}}_{\mbox{\scriptsize $k$}}}(x)
&=& \frac{1}{k!} \ln^k(x)~.
\end{eqnarray}
%----------------------------------------------------------------------------------------------
An example for relation (\ref{EQ1}) is 
%----------------------------------------------------------------------------------------------
\begin{eqnarray}
\label{EQ2}
S_{-2,1,1}(N) &=& -(-1)^N \Mvec\left[\frac{\HH_{0,1,1}(x)}{x+1}\right](N)
+ (-1)^N \zeta_3 \Mvec\left[\frac{1}{x+1}\right](N) \nonumber\\ && 
- \Li_4\left(\frac{1}{2} \right)
- \frac{\ln^4(2)}{24}
+ \frac{\ln^2(2) \zeta_2}{4}
- \frac{7 \ln(2) \zeta_3}{8}
+ \frac{\zeta_2^2}{8}~.
\end{eqnarray}
%----------------------------------------------------------------------------------------------
Here also special constants occur, either as infinite nested harmonic sums or as values of
the harmonic polylogarithms at $x=1$ as long as these are defined,
%----------------------------------------------------------------------------------------------
\begin{eqnarray}
\sigma_{\vec{a}} = \lim_{N \rightarrow \infty} S_{\vec{a}}(N)~. 
\end{eqnarray}
%----------------------------------------------------------------------------------------------
They are called multiple zeta values \cite{Blumlein:2009cf}.
Since for $N \rightarrow \infty$ the Mellin transforms in (\ref{EQ2}) vanish
one obtains
%----------------------------------------------------------------------------------------------
\begin{eqnarray}
\label{EQ2a}
\sigma_{-2,1,1}(N) &=& 
- \Li_4\left(\frac{1}{2} \right)
- \frac{\ln^4(2)}{24}
+ \frac{\ln^2(2) \zeta_2}{4}
- \frac{7 \ln(2) \zeta_3}{8}
+ \frac{\zeta_2^2}{8}~.
\end{eqnarray}
%----------------------------------------------------------------------------------------------
Here 
%----------------------------------------------------------------------------------------------
\begin{eqnarray}
\zeta_k = \sum_{l=1}^\infty \frac{1}{l^k},~~~k \geq 2,~~k \in \mathbb{N},~~~~\Li_k(x) = 
\sum_{l=1}^\infty
\frac{x^l}{l^k}
\end{eqnarray}
%----------------------------------------------------------------------------------------------
are the values of Riemann's $\zeta$-function and $\Li_k$ denotes the polylogarithm.
It is useful also to associate the symbol 
%----------------------------------------------------------------------------------------------
\begin{eqnarray}
\sigma_0 \equiv \sum_{k=1}^\infty \frac{1}{k}
\end{eqnarray}
%----------------------------------------------------------------------------------------------
to the multiple zeta values. All divergencies of the multiple zeta values can be expressed polynomially 
by $\sigma_0$.\footnote{This also holds for the special numbers occurring in case of the 
cyclotomic harmonic sums.}

Harmonic sums obey quasi-shuffle algebras \cite{HOFFMAN} as harmonic polylogarithms obey shuffle 
algebras.
These are implied by their multiplication relations at equal argument $N$ resp. $x$. The shuffle 
product 
is given by the sum of all permutations of indices of the two sets, which preserve the original ordering.
In case of the quasi-shuffle (stuffle \cite{Borwein:1999js})
algebras additional terms occur, cf.~\cite{Blumlein:2003gb}.
The product of two harmonic polylogarithms is thus given by
%----------------------------------------------------------------------------------------------
\begin{eqnarray}
\HH_a(x) \cdot \HH_{bcd}(x) =
\HH_a(x) \SH \HH_{bcd}(x) =
\HH_{abcd}(x)
+ \HH_{bacd}(x)
+ \HH_{bcad}(x)
+ \HH_{bcda}(x)~.
\end{eqnarray}
%----------------------------------------------------------------------------------------------
The stuffle and shuffle relations imply relations between the harmonic sums and harmonic polylogarithms,
respectively, which do not depend on their arguments $N$ and $x$ and are called algebraic relations
\cite{Blumlein:2003gb}. Both these algebras can also be applied to the multiple zeta values. Their action 
is not identical.

The algebraic relations are not the only relations of the harmonic sums or polylogarithms. Other
relations concern the argument of these quantities and are sometimes only valid for sub-classes
of indices. They are called structural relations. In case of harmonic sums they are implied by 
the duplication relation \cite{Blumlein:2009cf} and differentiation w.r.t. $N$ referring to 
an analytic continuation \cite{Blumlein:1998if,Blumlein:2009ta,Blumlein:2009fz,ABS13}.
At a given weight ${\sf w} = \sum_i |a_i|$ there are
%------------------------------------------------------------------------------------  
\begin{eqnarray}
 N_{\sf all}(\w)&=&2\cdot 3^{\w-1}
\end{eqnarray}  
%------------------------------------------------------------------------------------
harmonic sums. The algebraic $(A)$, differential $(D)$ and duplication $(H)$ relations
lead to
%------------------------------------------------------------------------------------
\begin{eqnarray}
 N_A(\w) = \frac{1}{\w}\sum_{d|\w}{\mu\left(\frac{{\w}}{d}\right)3^d},~~
 N_D({\w})= 4\cdot 3^{{\w}-2},~~
 N_H(\w) =  2\cdot 3^{\w-1}-2^{\w-1}, 
\end{eqnarray}
%------------------------------------------------------------------------------------
independent sums individually. Here, $\mu(\xi)$ denotes the M\"obius function~\cite{MOB}.
Applying these relations one obtains 
%------------------------------------------------------------------------------------
\begin{eqnarray}
 N_{ADH}(\w)&=&\frac{1}{\w}\sum_{d|\w}{\mu\left(\frac{\w}{d}\right)
\left[3^d-2^d\right]}-\frac{1}{\w-1}\sum_{d|\w-1}{\mu\left(\frac{\w-1}{d}\right)\left[3^d-2^d\right]}
\end{eqnarray}
%------------------------------------------------------------------------------------
independent sums. For $\w = 8$ the 4374 harmonic sums can thus be expressed by 486 basic sums.

The harmonic polylogarithms obey a general argument relation under the transformation
%------------------------------------------------------------------------------------
\begin{eqnarray}
x \rightarrow \frac{1-x}{1+x}~.
\end{eqnarray}
%------------------------------------------------------------------------------------
An example is
%------------------------------------------------------------------------------------
\begin{eqnarray}
\HH_{-1,0,1}\left[\frac{1-x}{1+x}\right] &=&
- \HH_{-1,1}(x) \left(\HH_0(x)+ \ln(2)\right)
+ \HH_{-1}(x) \left[\HH_{-1,1}(x) + \HH_{0,-1}(x) 
\right.
\nonumber\\ && \left.
+ \HH_{0,1}(x) 
- \zeta_2\right]
-2 \HH_{-1,-1,1}(x) - \HH_{0,-1,-1}(x) - \HH_{0,1,-1}(x) 
\nonumber\\ && 
- \frac{1}{2} \HH_{-1}^2(x) \left[\HH_0(x) 
+ \ln(2) \right]
+\frac{1}{6} \HH_{-1}^3(x) + \ln(2) \zeta_2
- \frac{5}{8} \zeta_3.
\end{eqnarray}
%------------------------------------------------------------------------------------
There are other relations for $x \rightarrow \{1-x, 1/x, x^2\}$ which are valid for special index 
sets \cite{Remiddi:1999ew}.

For the multiple zeta values over the alphabet $\{0,1,-1\}$ the shuffle and stuffle relations
imply all relations up to $\w =7$. From $\w =8$ the duplication and from $\w = 12$ the generalized 
duplication relation \cite{Blumlein:2009cf} leads to new relations. The number of newly contributing
basis elements for the lowest weights are, \cite{Blumlein:2009cf}, 
%------------------------------------------------------------------------------------
\begin{center}
\begin{tabular}{||l||c|c|c|c|c|c|c|c|c|c|c|c||}
\hline
\w        &    1 & 2 & 3 & 4 & 5 & 6 & 7 & 8 & 9 & 10 & 11 & 12 \\
\hline
\# basis  &    2 & 1 & 1 & 1 & 2 & 2 & 4 & 5 & 8 & 11 & 18 & 25 \\
\hline
\end{tabular}
\end{center}
%------------------------------------------------------------------------------------
I.e. up to \w = 12~~80 basis elements span the multiple zeta values.
Up to \w = 7 one possible representation reads \cite{Vermaseren:1998uu} 
%------------------------------------------------------------------------------------
\begin{eqnarray}
\left\{\sigma_0, \ln(2), \zeta_2, \zeta_3, \Li_4\left(\frac{1}{2}\right), \zeta_5, \Li_5\left(\frac{1}{2}\right),
\Li_6\left(\frac{1}{2}\right), \sigma_{-5,-1}, \zeta_7, \Li_7\left(\frac{1}{2}\right), \sigma_{-5,1,1}, 
\sigma_{5,-1,-1} \right\}.
\end{eqnarray}
%------------------------------------------------------------------------------------
It is not proven at present, whether these are all relations. For special sequences of harmonic sums
there are further relations, see e.g.~\cite{Blumlein:2009cf}. A global property of the MZVs over
the alphabet $\{0,1\}$, stating that they can be expressed in terms of MZVs having only indices 
$a_i = 2,3$, has been conjectured in \cite{HOFFC} and recently proven in \cite{BROWNH}.
%%%%%%%%%%%%%%%%%%%%%%%%%%%%%%%%%%%%%%%%%%%%%%%%%%%%%%%%%%%%%%%%%%%%%%%%%%%%%%%%%%%%%%%%%%%
\section{Cyclotomic Harmonic Sums, Polylogarithms, and Numbers}
%%%%%%%%%%%%%%%%%%%%%%%%%%%%%%%%%%%%%%%%%%%%%%%%%%%%%%%%%%%%%%%%%%%%%%%%%%%%%%%%%%%%%%%%%%%

\vspace*{1mm}
\noindent
The denominators $(x-1)$ and $(x+1)$ appearing in the harmonic polylogarithms form the first two 
cyclotomic polynomials \cite{LANG}. One may extend the alphabet in allowing  all cyclotomic
polynomials \cite{Ablinger:2011te}. They are given by
%------------------------------------------------------------------------------------
\begin{eqnarray}
\Phi_n(x) = \frac{x^n-1}{\prod_{d|n,d<n} \Phi_d(x)},~~~ d,n \in \mathbb{N} \backslash \{0\}.
\end{eqnarray}
%------------------------------------------------------------------------------------
We define the corresponding set of letters by
%------------------------------------------------------------------------------------
\begin{eqnarray}
f_0^0(x) = \frac{1}{x},~~~f_k^l(x) = \frac{x^l}{\Phi_k(x)}, k \in \mathbb{N} \backslash \{0\},~l \in 
\mathbb{N},~l < \varphi(k),
\end{eqnarray}
%------------------------------------------------------------------------------------
with $\varphi(k)$ being Euler's totient function \cite{LANG}. A few early examples of Mellin 
transforms
of cyclotomic polylogarithms were given in \cite{NIELSEN}. 

Iterating these letters one forms the cyclotomic polylogarithms which obey a shuffle algebra.
Applying the Mellin transform (\ref{EQmel}) one obtains combinations of the cyclotomic harmonic sums 
and the associated constants. As an extension of the normal harmonic sums, the single cyclotomic 
sums are given by  
%------------------------------------------------------------------------------------
\begin{eqnarray}
S_{l,m,n}(N)  = \sum_{k=0}^N \frac{({\rm sign}(n))^k}{(l k + m)^{|n|}}~,
\end{eqnarray}
%------------------------------------------------------------------------------------
i.e. harmonic sums with periodic gaps in the terms accounted. By iteration of this structure
the general cyclotomic sums are obtained. They occur in the calculation of massive Feynman 
diagrams. The cyclotomic sums obey algebraic and differentiation relations as well as three
multiple argument relations \cite{Ablinger:2011te}, for which counting relations are available.

The special constants being associated to the cyclotomic sums and polylogarithms extend
the multiple zeta values. The single sums at \w = 1 can be expressed by $\sigma_0, \ln(2)$ and $\pi$,
which replaces $\zeta_2 = \pi^2/6$ as a more fundamental constant. At higher cyclotomy $l$ also
the logarithms $\ln(3), \ln(\sqrt{2}-1), \ln(\sqrt{3}-1), \ln(\sqrt{5}-1)$ and several algebraic 
numbers occur. For $l \leq 6$ and $\w \geq 2$ the basic constants $\zeta_{2k+1}, \psi^{(2k+1)}(1/3), 
\Ti_{2k}(1), \psi^{(k)}(1/5), \psi^{(2k+1)}(2/5), \psi^{(k)}(1/8), \psi^{(2k)}(1/12)$ contribute. 
Here $\psi$ denotes the di-gamma function and $\Ti_{l}(1) = \sum_{k=0}^\infty (-1)^k/(2k+1)^l$, with 
$\Ti_2(1) = {\bf C}$ being Catalan's constant \cite{CATALAN}. These are the real representations of 
these 
constants. Likewise one may consider the infinite generalized harmonic sums with weights at the 
roots of unity
%------------------------------------------------------------------------------------
\begin{eqnarray} 
\lim_{N \rightarrow \infty} S_{k_1,...,k_m}(x_1,...,x_m;N)  \equiv 
\sigma_{k_1,...,k_m}(x_1,...,x_m),~~x_j \in {\cal C}_n,~n \geq 1,~k_1 \neq 1~\text{for}~x_1 = 1,  
\end{eqnarray}   
%------------------------------------------------------------------------------------
with ${\cal C}_n \in \{e_n|e_n^n = 1, e_n \in \mathbb{C}\}$. The real representations being discussed
above are related to these complex representations. For lower weights they have been studied
for cyclotomy $l \leq 20$ in \cite{Ablinger:2011te}.
%%%%%%%%%%%%%%%%%%%%%%%%%%%%%%%%%%%%%%%%%%%%%%%%%%%%%%%%%%%%%%%%%%%%%%%%%%%%%%%%%%%%%%%%%%%
\section{Generalized Harmonic Sums, Polylogarithms, and Numbers}
%%%%%%%%%%%%%%%%%%%%%%%%%%%%%%%%%%%%%%%%%%%%%%%%%%%%%%%%%%%%%%%%%%%%%%%%%%%%%%%%%%%%%%%%%%%

\vspace*{1mm}
\noindent
Generalized harmonic sums are defined by \cite{Moch:2001zr,Ablinger:2013cf}
%------------------------------------------------------------------------------------
\begin{eqnarray}
\label{EQ3}
S_{b,\vec{a}}(\zeta, \vec{\xi}; N) = \sum_{k = 1}^N \frac{\zeta^k}{k^b} S_{\vec{a}}(\vec{\xi};k),~~~b, a_i \in 
\mathbb{N} \backslash \{0\};~~ \zeta, \xi_i \in \mathbb{R} \backslash \{0\}.
\end{eqnarray}
%------------------------------------------------------------------------------------
The corresponding iterated integrals are built over the alphabet $\{0,\zeta, \vec{\xi}\}$.
To also associate the constants, i.e. the sums in the limit $N \rightarrow \infty$, one has to 
restrict the range of weights $\zeta, \xi_i$ accordingly to obtain convergent expressions. In 
intermediate physics results, however, divergent sums
for $|\xi_i| > 1$ do occur and have to be dealt with \cite{ABRSW13}. Known examples refer to alphabets
$\xi_i \in \{1,-1,1/2,-1/2,2,-2,1/3,-1/3,3,-3,...\}$.~\footnote{We would like to thank 
W.~Bernreuther and O.~Dekkers for a remark.} In some applications the weights $\xi_i$ are 
general real 
numbers. One may generalize the sums (\ref{EQ3}) introducing cyclotomic denominators \cite{Ablinger:2013cf}
%------------------------------------------------------------------------------------
\begin{eqnarray}
\label{EQ4}
\hspace*{-.5cm}
S_{\{a_1,b_1,c_1\}, ...,\{a_l,b_l,c_l\}}(s_1, ...,s_l; N)
&=& \sum_{k_1 = 1}^{N} \frac{s_1^k}{(a_1 k_1 + b_1)^{c_1}}
S_{\{a_2,b_2,c_2\}; ...;\{a_l,b_l,c_l\}}(s_2, ...,s_l; k_1), 
\end{eqnarray}
%----------------------------------------------------------------------------------------------
with $S_{\emptyset} = 1,~~a_i, c_i \in \mathbb{N} \backslash \{0\},~b_i
\in  \mathbb{N},~~s_i \in \mathbb{R} \backslash \{0\}, a_i > b_i.$ Also these sums are related to
the corresponding polylogarithms by the inverse Mellin transform. The elements of both spaces obey
(quasi)shuffle relations and a series of structural relations which were worked out in Ref.~\cite{Ablinger:2013cf}. 
An even wider class of special numbers is associated to the generalized (cyclotomic) harmonic sums 
and 
polylogarithms. A convenient way to work with these and the more special functions being listed above 
is provided by the {\tt Mathematica}-package {\tt HarmonicSums} \cite{Ablinger:2013hcp,Ablinger:2013cf}.
%%%%%%%%%%%%%%%%%%%%%%%%%%%%%%%%%%%%%%%%%%%%%%%%%%%%%%%%%%%%%%%%%%%%%%%%%%%%%%%%%%%%%%%%%%%
\section{Nested Binomial Sums}
%%%%%%%%%%%%%%%%%%%%%%%%%%%%%%%%%%%%%%%%%%%%%%%%%%%%%%%%%%%%%%%%%%%%%%%%%%%%%%%%%%%%%%%%%%%

\vspace*{1mm}
\noindent
In case of some Feynman diagrams \cite{ABRSW13} contributing to the massive Wilson coefficient 
for the structure function $F_2(x,Q^2)$ at higher scales of $Q^2$ at 3-loop order further extensions
of weighted generalized cyclotomic sums occur~: 
%----------------------------------------------------------------------------------------------
\begin{eqnarray}
&& \sum_{i=1}^N \binom{2i}{i}(-2)^i \sum_{j=1}^i \frac{1}{\ds j \binom{2j}{j}}
S_{1,2}\left(\tfrac{1}{2},-1;j\right) \\
&&  = \int_0^1 dx \frac{x^N-1}{x-1} \sqrt{\frac{x}{8+x}}\left[{\rm H}^*_{w_{17},-1,0}(x) 
- 2
{\rm H}^*_{w_{18},-1,0}(x)\right]
\nonumber\\ &&
+ \frac{\zeta_2}{2} \int_0^1 dx \frac{(-x)^N-1}{x+1} \sqrt{\frac{x}{8+x}}\left[{\rm
H}^*_{12}(x)
-2 {\rm H}^*_{13}(x)\right]
%\nonumber\\ &&
+ c_3 \int_0^1 dx \frac{(-8x)^N-1}{x+\frac{1}{8}} \sqrt{\frac{x}{1-x}}~,
\nonumber
\end{eqnarray}
%----------------------------------------------------------------------------------------------
where
%----------------------------------------------------------------------------------------------
\begin{eqnarray}
w_{12} = \frac{1}{\sqrt{x(8-x)}},~~
&& w_{13} = \frac{1}{(2-x)\sqrt{x(8-x)}},\nonumber\\
w_{17} = \frac{1}{\sqrt{x(8+x)}},~~
&&w_{18} = \frac{1}{(2+x)\sqrt{x(8+x)}}~.
\label{eq:BINS1}
\end{eqnarray}
%----------------------------------------------------------------------------------------------
The iterated integrals $\HH^*$ are defined here on the interval $[x,1]$.
The new element consists in binomial $\binom{2i}{i}$ terms emerging both in the numerators and denominators
of the finite nested sums.\footnote{Infinite single sums with binomial weights in the numerator and denominator
over (generalized) harmonic sums were studied in \cite{Davydychev:2003mv,Weinzierl:2004bn}.}
About $100$ independent nested sums of similar type contribute. The associated iterated 
integrals request square-root valued alphabets with about 30 new letters, extending those in case of 
the
generalized harmonic polylogarithms. Examples are given in  Eqs.~(\ref{eq:BINS1}).
%%%%%%%%%%%%%%%%%%%%%%%%%%%%%%%%%%%%%%%%%%%%%%%%%%%%%%%%%%%%%%%%%%%%%%%%%%%%%%%%%%%%%%%%%%%
\section{Elliptic Integrals}
%%%%%%%%%%%%%%%%%%%%%%%%%%%%%%%%%%%%%%%%%%%%%%%%%%%%%%%%%%%%%%%%%%%%%%%%%%%%%%%%%%%%%%%%%%%

\vspace*{1mm}
\noindent
Nested binomial sums, weighting generalized cyclotomic sums, lead to square-root values 
letters. One may now imagine that Mellin convolutions of these quantities do also 
emerge, as it happens already for ordinary harmonic polylogarithms. Let us consider a
simple case of this kind~: 
%--------------------------------------------------------------------------------------------$
\begin{eqnarray}
T(x) &=& \frac{1}{\sqrt{1-x}} \otimes
  \frac{1}{\sqrt{1-x}} 
= \int_x^1 \frac{dy}{y} \frac{1}{\sqrt{1-y}} \frac{1}{\sqrt{1-\tfrac{x}{y}}}
\nonumber\\
&=& 2i \left[{\bf F}\left(\arcsin\left(\frac{1}{\sqrt{x}}\right),x\right) - {\bf K}(x)\right]
\end{eqnarray}
%--------------------------------------------------------------------------------------------$
It involves the elliptic integrals
%--------------------------------------------------------------------------------------------$
\begin{eqnarray}
F(x;k) &=& \int_0^x \frac{dt}{\sqrt{(1-t^2)(1-k^2t^2)}}\\
K(k)   &=& F(1,k) = \frac{\pi}{2} {_2F_1}\left(\frac{1}{2}, \frac{1}{2}; 1; k^2\right)~.
\end{eqnarray}
%--------------------------------------------------------------------------------------------$
On the other hand, the Mellin transform of $T(x)$ yields the following simple expression.
%--------------------------------------------------------------------------------------------$
\begin{eqnarray}
\Mvec[T(x)](N) = \int_0^1 dx x^N T(x) = \frac{4^{2N}}{\displaystyle \binom{2N}{N}^2 
(N+\tfrac{1}{2})^2}~.
\end{eqnarray}
%--------------------------------------------------------------------------------------------$
Higher powers of the binomial $\binom{2N}{N}$ emerge in Mellin space, which are 
seemingly one source of elliptic integrals in $x$-space. One has to contest, that the
$N$-space expression is more simple here. 
%%%%%%%%%%%%%%%%%%%%%%%%%%%%%%%%%%%%%%%%%%%%%%%%%%%%%%%%%%%%%%%%%%%%%%%%%%%%%%%%%%%%%%%%%%%
\section{Analytic Continuations the various Sums}
%%%%%%%%%%%%%%%%%%%%%%%%%%%%%%%%%%%%%%%%%%%%%%%%%%%%%%%%%%%%%%%%%%%%%%%%%%%%%%%%%%%%%%%%%%%

\vspace*{1mm}
\noindent
In Mellin space one may thoroughly perform the solution of the QCD evolution equations
analytically, cf. e.g. \cite{Blumlein:1997em}. The convolution of the Wilson coefficients
with the evolved parton distribution functions is given by a simple product. This representation 
is therefore preferred in fitting the non-perturbative parton distribution functions. On 
the numerical side only one final contour integral around the singularities of the problem has to be 
performed.
This requests the analytic continuation of the variable $N$ in the nested sums to $N \in \mathbb{C}$.
Observing the crossing relations of the respective process the analytic continuation proceeds either
from the even {\it or} the odd values of $N$. First the singularities in the complex plane have to 
be determined. 
For the harmonic sums and cyclotomic harmonic sums the singularities are located at the non-positive integers 
and one obtains meromorphic functions. This is not necessarily the case for generalized sums since they may 
diverge in some cases exponentially as $N \rightarrow \infty$. Whenever an asymptotic expansion exists, it can 
be calculated analytically \cite{Ablinger:2013cf,Ablinger:2013hcp,ABS13}\footnote{Precise numerical 
implementations for the analytic continuations of special Mellin-transforms up to those needed
to express the 3-loop anomalous dimensions were given in \cite{Blumlein:2000hw,Blumlein:2005jg}.}
and thus be given at arbitrary 
precision in principle. Starting with this representation, the shift-property of the nested sums for $N 
\rightarrow N+1$ allows to arrive at any non-singular point in the complex plane using a thoroughly analytic 
representation to be evaluated numerically. Physical quantities like the massless and the 
known massive Wilson coefficients and massive operator matrix elements
\cite{ANDIM31,ANDIM32,Ablinger:2010ty,Blumlein:2012vq,GQ} possess regular asymptotic 
representations. For these 
quantities a corresponding representation is therefore possible. This also applies for the Wilson coefficients 
of the Drell-Yan process, hadronic Higgs-boson production \cite{TWOLOOP1} and time-like quantities 
\cite{TWOLOOP2}. Furthermore, precise representations can be derived also in case cross sections are given 
numerically only, cf.~\cite{Alekhin:2003ev}.
%%%%%%%%%%%%%%%%%%%%%%%%%%%%%%%%%%%%%%%%%%%%%%%%%%%%%%%%%%%%%%%%%%%%%%%%%%%%%%%%%%%%%%%%%%%
\section{Conclusions}
%%%%%%%%%%%%%%%%%%%%%%%%%%%%%%%%%%%%%%%%%%%%%%%%%%%%%%%%%%%%%%%%%%%%%%%%%%%%%%%%%%%%%%%%%%%

\vspace*{1mm}
\noindent
The mathematical functions expressing Feynman diagrams in $N$-space form a hierarchy starting 
with rational functions, harmonic sums, followed by generalized harmonic sums, cyclotomic sums, 
their generalization, binomially weighted generalized cyclotomic sums, etc. Accordingly, the 
corresponding iterated integrals and special numbers are organized. The relations of the different
quantities can be illustrated by Figure~1 \cite{Ablinger:2013jta}.

%\newpage

\def\circlea{(-8.3cm,6cm) circle (1.8cm and 1cm)}
\def\circleb{(-6.5cm,6cm) circle (1.8cm and 1cm)}
\def\circlec{(-1.8cm,6cm) circle (1.8cm and 1cm)}
\def\circled{(-0.0cm,6cm) circle (1.8cm and 1cm)}
\def\circlee{(-8.3cm,2cm) circle (1.8cm and 1cm)}
\def\circlef{(-6.5cm,2cm) circle (1.8cm and 1cm)}
\def\circleg{(-1.8cm,2cm) circle (1.8cm and 1cm)}
\def\circleh{(-0.0cm,2cm) circle (1.8cm and 1cm)}

\tikzset{filled/.style={fill=circle area, draw=circle edge, thick}, outline/.style={draw=circle edge, thick}}

\colorlet{circle edge}{black!100}
\colorlet{circle area}{black!20}
\setlength{\parskip}{5mm}
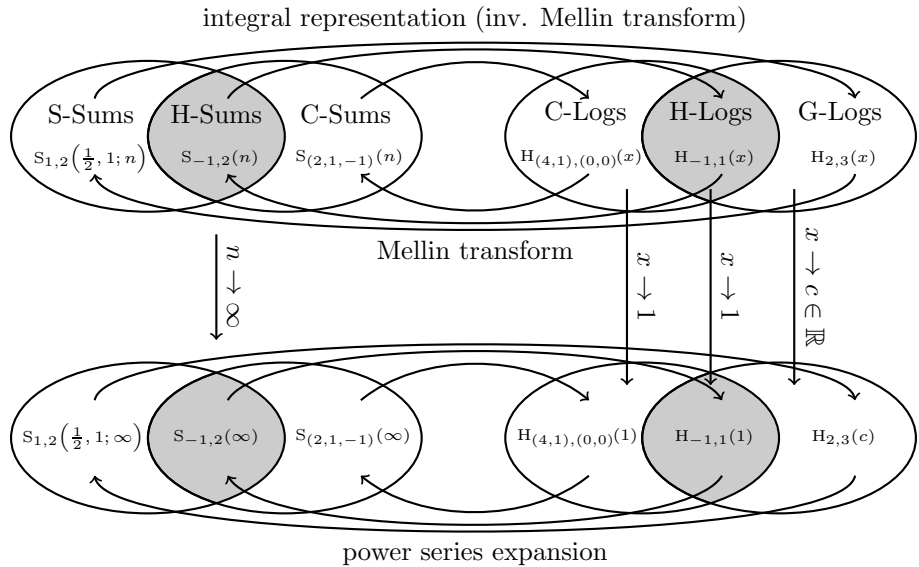
\begin{figure}[H]
\centering
\begin{tikzpicture}
\centering
     \begin{scope}
         \clip \circlea;
         \fill[filled] \circleb;
     \end{scope}
     \draw[outline] \circlea;
     \draw[outline] \circleb;
     \draw (-7.40cm,6.3cm) node[font=\small] {H-Sums};
     \draw (-7.35cm,5.7cm) node[font=\tiny] {$\S{-1,2}n$};
     \draw (-9.05cm,6.3cm) node[font=\small] {S-Sums};
     \draw (-9.05cm,5.7cm) node[font=\tiny] {$\S{1,2}{\frac{1}{2},1;n}$};
     \draw (-5.70cm,6.3cm) node[font=\small] {C-Sums};
     \draw (-5.65cm,5.7cm) node[font=\tiny] {$\S{(2,1,-1)}n$};
     \begin{scope}
         \clip \circlec;
         \fill[filled] \circled;
     \end{scope}
     \draw[outline] \circlec;
     \draw[outline] \circled;
     \draw (-0.90cm,6.3cm) node[font=\small] {H-Logs};
     \draw (-0.85cm,5.7cm) node[font=\tiny] {$\H{-1,1}x$};
     \draw (-2.55cm,6.3cm) node[font=\small] {C-Logs};
     \draw (-2.60cm,5.7cm) node[font=\tiny] {$\textnormal{H}_{(4,1),(0,0)}\hspace{-0.05cm}(x)$};
     \draw (+0.80cm,6.3cm) node[font=\small] {G-Logs};
     \draw (+0.85cm,5.7cm) node[font=\tiny] {$\H{2,3}x$};

     \draw[thick,->] (-9cm,6.5cm) .. controls +(80:1cm) and +(100:1cm) .. (1cm,6.5cm) node[midway,sloped,above,font=\small] {integral representation (inv. Mellin transform)};
     \draw[thick,->] (-7.25cm,6.5cm) .. controls +(60:1cm) and +(120:1cm) .. (-0.75cm,6.5cm);
     \draw[thick,->] (-5.50cm,6.5cm) .. controls +(40:1cm) and +(140:1cm) .. (-2.5cm,6.5cm);

     \draw[thick,<-] (-9cm,5.5cm) .. controls +(280:1cm) and +(260:1cm) .. (1cm,5.5cm) node[midway,sloped,below,font=\small] {Mellin transform};
     \draw[thick,<-] (-7.25cm,5.5cm) .. controls +(300:1cm) and +(240:1cm) .. (-0.75cm,5.5cm);
     \draw[thick,<-] (-5.5cm,5.5cm) .. controls +(320:1cm) and +(220:1cm) .. (-2.5cm,5.5cm);

     \begin{scope}
         \clip \circlee;
         \fill[filled] \circlef;
     \end{scope}
     \draw[outline] \circlee;
     \draw[outline] \circlef;
     \draw (-7.40cm,2cm) node[font=\tiny] {$\S{-1,2}\infty$};
     \draw (-9.15cm,2cm) node[font=\tiny] {$\S{1,2}{\frac{1}{2},1;\infty}$};
     \draw (-5.60cm,2cm) node[font=\tiny] {$\S{(2,1,-1)}\infty$};
     \draw[thick,->] (-7.4cm,4.7cm) -- (-7.4cm,3.3cm) node[midway,sloped,above,font=\small] {$n\rightarrow\infty$};

     \begin{scope}
         \clip \circleg;
         \fill[filled] \circleh;
     \end{scope}
     \draw[outline] \circleg;
     \draw[outline] \circleh;
     \draw (-0.85cm,2cm) node[font=\tiny] {$\H{-1,1}1$};
     \draw (-2.65cm,2cm) node[font=\tiny] {$\textnormal{H}_{(4,1),(0,0)}\hspace{-0.05cm}(1)$};
     \draw (0.85cm,2cm) node[font=\tiny] {$\H{2,3}c$};
     \draw[thick,->] (-0.9cm,5.3cm) -- (-0.9cm,2.7cm) node[midway,sloped,above,font=\small] {$x\rightarrow 1$};
     \draw[thick,->] (-2.0cm,5.3cm) -- (-2.0cm,2.7cm) node[midway,sloped,above,font=\small] {$x\rightarrow 1$};
     \draw[thick,->] (0.2cm,5.3cm) -- (0.2cm,2.7cm) node[midway,sloped,above,font=\small] {$x\rightarrow c\in \mathbb R$};

     \draw[thick,<-] (-9cm,1.5cm) .. controls +(280:1cm) and +(260:1cm) .. (1cm,1.5cm) node[midway,sloped,below,font=\small] {power series expansion};
     \draw[thick,<-] (-7.25cm,1.5cm) .. controls +(300:1cm) and +(240:1cm) .. (-0.75cm,1.5cm);
     \draw[thick,<-] (-5.5cm,1.5cm) .. controls +(320:1cm) and +(220:1cm) .. (-2.5cm,1.5cm);

     \draw[thick,->] (-9cm,2.5cm) .. controls +(80:1cm) and +(100:1cm) .. (1cm,2.5cm);
     \draw[thick,->] (-7.25cm,2.5cm) .. controls +(60:1cm) and +(120:1cm) .. (-0.75cm,2.5cm);
     \draw[thick,->] (-5.50cm,2.5cm) .. controls +(40:1cm) and +(140:1cm) .. (-2.5cm,2.5cm);

\end{tikzpicture}
\caption{\label{Ifig1}\footnotesize Connection between harmonic sums (H-Sums), S-sums (S-Sums) and cyclotomic harmonic 
sums (C-Sums), 
their values at infinity and harmonic polylogarithms (H-Logs), generalized harmonic polylogarithms (G-Logs) and 
cyclotomic harmonic polylogarithms (C-Logs) and their values at special constants.}
\end{figure}

\noindent
The cyclotomic polynomials 
provide a natural extensions of the letters used with iterated integrals leading to harmonic 
polylogarithms. Corresponding terms occur in massive higher order calculations. The Mellin 
transform associates the nested sums and the iterated integrals. Both classes form quasi-shuffle 
resp. shuffle algebras and obey structural resp. argument-induced relations. Similar relations hold
for the different sets of special numbers. One expects an even richer structure in case 
of multi-leg integrals at higher loop orders, a territory which is  widely unexplored still. In this 
way Feynman diagrams generate a still growing number of new classes of mathematical structures. 
Knowing their relations greatly helps to simplify the theoretical calculations and also allows 
for better numerical representations.

\vspace*{2mm}\noindent
{\bf Acknowledgment.}~We would like to thank A.~De Freitas, C.~Raab,
and J.~Vermaseren for discussions. This work has been supported in part by DFG
Sonderforschungsbereich Transregio 9, Computergest\"utzte Theoretische
Teilchenphysik, by the Austrian Science Fund (FWF) grant P20347-N18, SFB F50 (F5009-N15)
and by the EU Network {LHCPHENOnet}~PITN-GA-2010-264564.

\medskip
\section*{References}
%%%%%%%%%%%%%%%%%%%%%%%%%%%%%%%%%%%%%%%%%%%%%%%%%%%%%%%%%%%%%%%%%%%%%%%%%%%%%%%%%%%%%%%%%%%%%%%

\end{document}